\documentclass[aps,pra,twocolumn,superscriptaddress,longbibliography]{revtex4-2}  

\usepackage{stmaryrd}
\usepackage{physics}
\usepackage{graphicx}
\usepackage{color}
\usepackage{bm,bbold}
\usepackage{bbm}
\usepackage{enumerate}
\usepackage{float}
\usepackage{comment}
\usepackage[caption=false]{subfig}
\graphicspath{figures} % Location of the graphics files
\usepackage{amsfonts, amsmath, amsthm, amssymb}

\usepackage[colorlinks,bookmarks=false,citecolor=blue,linkcolor=red,urlcolor=blue]{hyperref}

\usepackage[commandnameprefix=always]{changes}

\begin{document}

\title[ATI]{Spatial Leggett-Garg inequalities}

\author{Guillem Müller-Rigat}
\affiliation{ICFO-Institut de Ciencies Fotoniques, The Barcelona Institute of Science and Technology, Castelldefels (Barcelona) 08860, Spain}
\author{Donato Farina}
\affiliation{Physics Department E. Pancini, Università degli Studi di Napoli Federico II, Complesso Universitario Monte S. Angelo, I-80126 Napoli, Italy.
}
\author{Maciej Lewenstein}
\affiliation{ICFO-Institut de Ciencies Fotoniques, The Barcelona Institute of Science and Technology, Castelldefels (Barcelona) 08860, Spain}
\affiliation{ICREA, Pg. Lluís Companys 23, 08010 Barcelona, Spain}
\author{Andrea Tononi}
\affiliation{ICFO-Institut de Ciencies Fotoniques, The Barcelona Institute of Science and Technology, Castelldefels (Barcelona) 08860, Spain}
\affiliation{Department de F\'isica, Universitat Polit\`ecnica de Catalunya, Campus Nord B4-B5, E-08034, Barcelona, Spain}
\date{\today}

\begin{abstract}
We formulate a spatial extension of the Leggett-Garg inequality by considering three distant observers locally measuring a many-body system at three subsequent times. The spatial form is especially suitable to test the ability of quantum devices to generate spatio-temporal correlations in a Hamiltonian-agnostic manner. We illustrate our proposal for a Heisenberg chain in a magnetic field, showing indeed that the first inequality-violation time scales proportionally to the distance between measuring parties. We attribute this phenomenon to Lieb-Robinson physics and, confirming this connection, we find that violations are anticipated when increasing the interaction range. The inequality violation is readily observable in current quantum simulation platforms, particularly Rydberg atoms arrays and ultracold atoms in optical lattices. In outlook, spatial Leggett-Garg inequalities constitute a practical tool for benchmarking the complex non-relativistic dynamics of many-body quantum systems.
\end{abstract}

\maketitle  

\section{Introduction}
The Leggett-Garg inequality (LGI) \cite{leggettgarg1985} provides a testable criterion to analyze the correlations in time of an evolving physical system.
Its formulation assumes intuitive classical concepts, like macroscopic realism, noninvasive measurability and induction \cite{emary2014, vitagliano2023}, yielding constraints on the statistics of observations at multiple times. 
The first Leggett-Garg inequality violation by quantum systems was observed experimentally in 2010 \cite{palacios2010experimental}.
This evidence, together with a long series of subsequent experiments \cite{goggin2011, dressel2011, athalye2011, knee2012, zhou2015, formaggio2016, knee2016, katiyar2017, wang2017, liu2019, wang2020, santini2022, zhan2023, kreuzgruber2024, Lu2024, Cheng2025, Chatterjee2025} (see also the reviews \cite{emary2014, vitagliano2023}), requires us to adopt quantum mechanics as the everyday working tool for correctly describing correlations in microscopic systems.

In the past decades, Leggett-Garg inequalities have been extended to various scenarios, including multiple measurements and times \cite{avis2010, emary2014, budroni2013bounding}, macroscopic spin-$j$ systems (with $j$ large) \cite{kofler2008}, and neutrino oscillations \cite{formaggio2016, naikoo2019, blasone2023},
and its loopholes \cite{wilde2012addressing, budroni2015quantum} and limitations \cite{majidy2021, halliwell2017} have been addressed.
In spin chains, extended Leggett-Garg inequalities based on repeated local or collective measurements have been proposed to witness equilibrium phase transitions~\cite{GmezRuiz2016, mendoza2019enhancing} and as a quantum coherence witness \cite{santini2022}.
Spatial versions, however, in which observers at different locations measure distinct parts of a time-evolving many-body system, have never been proposed. 
Such an extension could enable the analysis and certification of non-equilibrium features of many-body dynamics, including the light cone spreading of correlations~\cite{Cubitt2008, Cheneau2012, Bonnes2014}.

In this paper, we introduce a spatial Leggett-Garg inequality for a protocol in which multiple parties perform spatially separated sequential measurements (see Fig.~\ref{fig-scheme}).
In general, and in contrast with the standard Leggett-Garg inequality, the violation of the spatial Leggett-Garg inequality requires an interacting Hamiltonian.
This feature allows its use as a figure of merit to benchmark the ability of a quantum system to generate spatio-temporal correlations in a system-agnostic way.

In principle, our conceptual extension is applicable to physical systems in any dimension and governed by arbitrary Hamiltonians.
However, to illustrate the concepts in an experimentally relevant setting, we analyze the Leggett-Garg correlator for a (spatiotemporally) symmetric measurement protocol performed on a spin chain evolving under the Heisenberg Hamiltonian in a magnetic field \cite{arnesen2001natural, amico2008entanglement}. 
We observe the reduction of the Leggett-Garg inequality violation when increasing the system size, showing that the propagation of a quantum signal through the chain is increasingly difficult for larger distances. Along this line, since time is necessary to establish correlations, the first instance of violation takes place at times increasing with the system size. 
This behavior is coherent with the fact that quantum correlations can be distributed in spin chains with finite velocity \cite{Cheneau2012}, theoretically limited by the Lieb-Robinson bound~\cite{lieb1972}, that we can practically estimate. 
Furthermore, we show through our method that, compared to other models with similar Lieb-Robinson velocities, the maximal violation of the spatial Leggett-Garg inequality is actually achieved by the Heisenberg Hamiltonian. 
This result provides a method to certify its realization in quantum devices in an implementation-agnostic manner.

A further practical purpose of our work is providing a concrete proposal for the experimental observation of spatial Leggett-Garg inequality violations in spin chains. 
Suitable platforms are, e.g., Rydberg atoms \cite{browaeys2020many, PRXQuantum.3.020303, geier2021floquet, miller2024two}, trapped atoms \cite{bloch2012quantum, gross2017, doi:10.1126/science.abk2397}, ions \cite{blatt2012quantum, RevModPhys.93.025001}, and ultracold atoms \cite{jepsen2020spin, PhysRevX.11.011011}. Performing a spatial Leggett-Garg experiment will allow quantifying, in a platform-independent manner, the ability to generate and propagate spatiotemporal correlations in such implementations. 

\section{Spatial Leggett-Garg inequalities}

\begin{figure}[h]
\centering
\includegraphics[width=.95\linewidth]{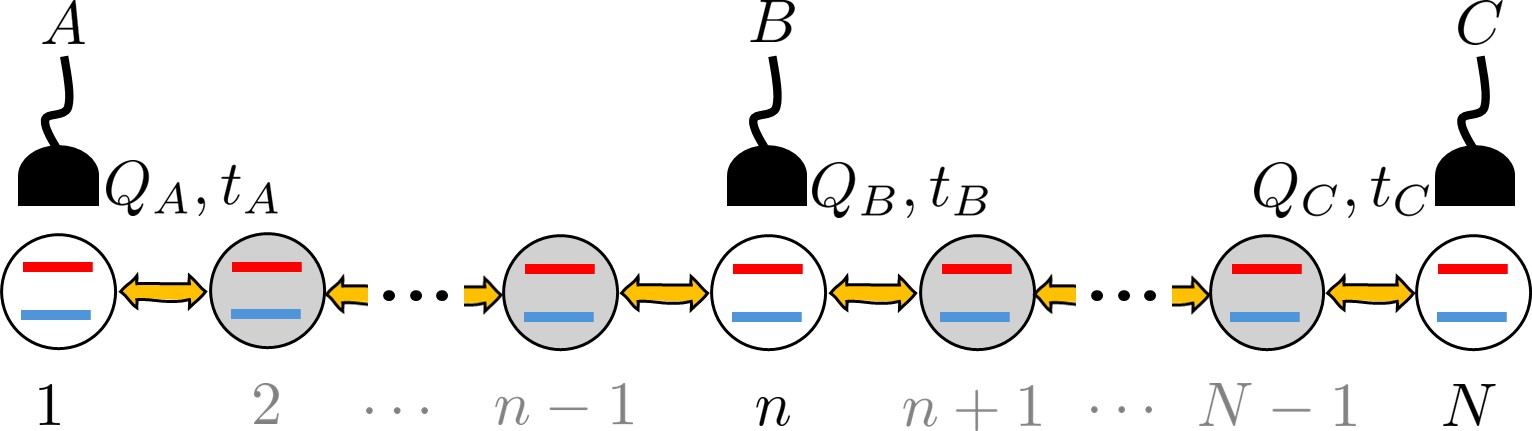}
\caption{
Measurement setup of the spatial Leggett-Garg inequality.
The three parties $A$, $B$, and $C$, displaced in space and connected by spin-$1/2$ chains, measure $Q_A$, $Q_B$, and $Q_C$ at the respective times $t_A$, $t_B$, and $t_C$.
In our implementation of the inequality we consider the spatiotemporally symmetric scenario $N = 2n-1$, $t_A = 0$, $t_B = t$, and $t_C = 2t$.}
\label{fig-scheme}
\end{figure}

The original formulation of the Leggett-Garg inequality considers an observer performing the binary measurement $Q$ on a system at the three subsequent times: $t_A$, $t_B$ and $t_C$. Next, one introduces the two-times correlation functions $c_{X,Y} =\langle Q(t_X) Q(t_Y) \rangle = \sum_{\{q_X,q_Y\} = \pm 1} q_X q_Y \, P_{q_Xq_Y}(Q,t_X,t_Y)$, 
with $P_{xy}(Q,t_X,t_Y)$ being the joint probabilities of measuring the dichotomic observable $Q$ at time $t_X$ and getting $q_X$, and again $Q$ at time $t_Y$ and getting $q_Y$. It can be shown that the quantity $K = c_{A,B} + c_{B,C} - c_{A,C}$ satisfies the Leggett-Garg inequality $-3 \leq K\leq 1$ for a system respecting the macrorealistic assumptions, namely, macrorealism \textit{per se} (i.e., that the observable $Q$ has a well-defined value at each time $t$), noninvasive measurability, and induction (i.e. the fact that the outcome of a measurement is not influenced by what will be measured at later times)~\cite{vitagliano2023}. The upper-bound value $K=1$, in particular, can be derived by considering the combinations that the value of $Q$ takes in the possible classical scenarios and evaluating the correlators \cite{emary2014}.

The Leggett-Garg inequality is violated by quantum dynamics, as can be seen already in simple two-level single-particle systems. 
A clear example is a spin-$1/2$ initialized in the state $\ket{+}$, with $\sigma^x \ket{+} = \ket{+}$ and $\sigma^{x,y,z}$ the Pauli matrices, and evolving under the Hamiltonian $\hat{H} = -h \sigma^z/2$. 
For three $\sigma^x$ measurements at times $t_A = 0$, $t_B = t$, and $t_C = 2t$, one obtains $K = 2 \cos(h t) - \cos (2h t)$, which yields the maximal inequality violation $K=3/2$ at times $t = \pi/(3h) + 2p\pi/h$ and $t = 5\pi/(3h) + 2p\pi/h$, where $p$ is a generic integer  \cite{kofler2008}.

Let us now consider a many-body system to conceptually extend the Leggett-Garg inequality in space.
For concreteness, see Fig.~\ref{fig-scheme}, we analyze a one-dimensional spin chain made of $N$ two-level systems.
We consider the three observers $A$, $B$, and $C$, each one measuring the same observable at their respective chain sites $1$, $n$, and $N$, respectively at times $t_{A}$, $t_{B}$, and $t_{C}$.
For simplicity, we will analyze the spatially symmetric configuration corresponding to $N = 2n-1$ and simply take $t_A = 0$, $t_B = t$, and $t_C = 2t$.
Note that, in this way, $n-1$ labels the lattice distance between the measuring parties: for $n=1$ all parties coincide; then $n=2$ corresponds to three adjacent parties, $n=3$ corresponds to three parties with one spin in between, and so on.
It is easy to generalize the above simplifications, as well as to relax the assumptions on the spatial dimension, boundary conditions, chain length, and number of levels.

To derive the spatial Leggett-Garg inequality, we delocalize the measurements in space and redefine the correlation functions as $C_{X,Y} = \langle Q_{X}(t_X) Q_Y(t_Y) \rangle$, which describes the sequential expectation value of measuring $Q$ at site $X$ and time $t_X$ and later $Q$ at site $Y$ and time $t_Y$. We define the spatial extension of the Leggett-Garg inequality as 
\begin{equation}
-3 \leq K_n := C_{1,n} + C_{n,2n-1} - C_{1,2n-1}\leq 1.
\label{LGspatial}
\end{equation}

Let us now put Eq.~\eqref{LGspatial} into context. 
For $n=1$, it represents the ordinary LGI, reviewed above for a single spin and whose violation has been detected in several systems over the years (among all, quarks~\cite{Cheng2025}, qutrits~\cite{wang2017}, trapped ion qubits~\cite{Lu2024}, and molecules~\cite{Chatterjee2025}).
The $n>1$ extension adds intrinsically different features to this standard LGI scenario.
First, since we consider a multipartite system rather than a single system, our inequality captures the correlation dynamics \textit{within} the system parts. 
Second, the $n>1$ violation would indicate the incompatibility with a macrorealistic model for the global many-body system. 
Third, the correlator Eq.~\eqref{LGspatial} is easier to implement than other well established tools to diagnose quantum dynamics such as out-of-time-ordered correlations~\cite{GarcaMata2023, lin2018, riddell2019}, which typically require reversing the system's dynamics.
Finally, note that various experiments over the years have aimed to demonstrate macrorealism by protecting the system against classically invasive measurements using techniques ranging from quantum memories~\cite{Paz1993} to quantum non-demolition measurements~\cite{budroni2015quantum}, always remaining in the \textit{single-system framework} akin to $n=1$. 

\section{Applications}

Our formulation Eq.~\eqref{LGspatial} does not rely on the calibration of the measurement setting $Q$, the time evolution, or the initial state. 
Let us however specify these choices to illustrate the quantum dynamics of $K_n$ in a concrete setting.
We consider a Heisenberg chain in a magnetic field with open boundaries: 
\begin{equation}
\begin{split}
H =& \sum_{i=1}^{N-1}\left( J_x\sigma_i^{x} \sigma_{i+1}^{x} + J_y\sigma_i^{y} \sigma_{i+1}^{y} +J_z\sigma_i^{z} \sigma_{i+1}^{z}  \right) - \frac{h}{2} \sum_{i=1}^{N} %\left( 
\sigma^z_i, %+ \mathbbm{1}_i 
%\right), 
\end{split}
\label{XXHamiltonian}
\end{equation}
where $J_x$, $J_y$, $J_z$, and $h$ are the coupling constants. We first consider the isotropic case $J_x=J_y=J_z := J$. 
As initial state, we take the product state $\ket{\Psi} =  \ket{+}^{\otimes N}$. Finally, the quantum correlators involved in the inequality are then readily computed as $C_{X,Y} = \Re{\bra{\Psi}Q_X(t_X)Q_Y(t_Y)\ket{\Psi}}$~\cite{Fritz2010}, with $Q_X = \sigma_X^x$.

\begin{figure}[t!]
%\centering
\includegraphics[width=.99\linewidth]{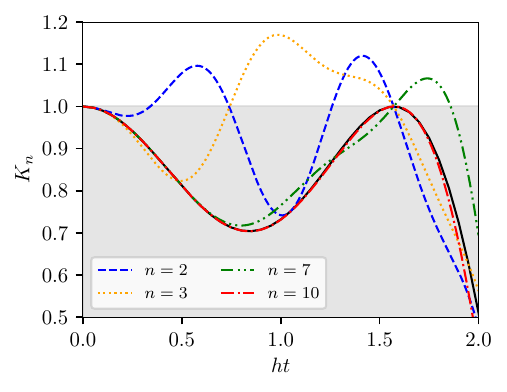}

\caption{Dynamics of the spatial Leggett-Garg inequality Eq.~\eqref{LGspatial} in the spin chain Eq.~\eqref{XXHamiltonian} with $J=1$ and $h =1$ for various distances $n$ between measuring parties and $Q = \sigma^x$. 
The black solid curve represents the noninteracting case of $J=0$ and $h=1$, in which no correlations can be established and the inequality is never violated. Note that violations are more difficult in longer chains, which tend to the noninteracting curve, and that the first violation time is ordered with $n$ (see Fig.~\ref{fig3}). 
Note that, for $n\geq 10$, the first violation time moves to the next period, i.e. the region around $ht \sim 2 \pi$ of Fig.~\ref{fig4}, where inequality violation revivals for $n < 10$ also occur.  
}
\label{LG_spatial}
\end{figure}

The value of $K_n$ can be calculated analytically in the simple case of $J=0$. 
In this case, the result does not depend on $n$ since all spins evolve independently under the local Hamiltonian $\propto \sigma^z$, and thus rotate in the $xy$ plane under the action of the external field $h$. 
If the parties respectively measure the observables $\sigma_1^x$, $\sigma_{n}^x (t)$, and $\sigma_{2n-1}^x (2t)$, where the lower index indicates the chain site of the measurement, we obtain $K_n = \left[ 3 \cos(h t) - 2 \cos(2h t) + \cos(3h t)  \right]/2$ (see Appendix~\ref{sec:app_noniteracting}). This function, depicted as the black solid line in Fig.~\ref{LG_spatial}, does not lead to any inequality violation. 
This result suggests that, to observe LGI violations, the interactions between the spins are necessary. Indeed, for any non-interacting Hamiltonian one finds $[Q_X(t_X),Q_Y(t_Y)] = 0$  for $X\neq Y$ and $X, Y\in \{A,B,C\}$, and, under this assumption, only one can prove that the inequality Eq.~\eqref{LGspatial} cannot be violated (see Appendix). However, not all interacting evolutions might be detected by the inequality.

Let us now consider nonzero interactions and analyze the violations of the spatial LGI.
We discuss the numerical results for the Hamiltonian \eqref{XXHamiltonian}, presented in Fig.~\ref{LG_spatial} for the values of $J=1$ and of $h = 1$.
Setting $J \neq 0$ allows for violations of the inequality, which occur at times close to the maxima of the noninteracting black curve, which we are perturbing by considering $J>0$.
Also note that, as $n$ increases, the short-time noninteracting profile is recovered for increasingly large time intervals.
Thus, the overall message of Fig.~\ref{LG_spatial} is that, if there are violations for a certain initial state, Hamiltonian and measurements, then increasing the length of the chain tends to increase the first violation time.
This phenomenon is a consequence of the existence of a maximum velocity, the Lieb-Robinson bound \cite{lieb1972}, at which correlations can propagate in a spin chain.  
We actually note that the Lieb-Robinson bound can be exceeded (and the LGI inequality violated) by small correlations. 
However, since these are exponentially suppressed with the system size, our formalization can characterize the non-relativistic correlation dynamics of the system quantitatively.

Additionally, for a given Hamiltonian and initial state, the measurements may be more or less optimal for observing a violation at a given time.
To reduce this variability, we perform a measurement optimization to maximize the value of $K_n$ at each time step \cite{budroni2013bounding}. 
The details of such optimization are discussed in Appendix~\ref{sec:app_optimization}, where we indeed verify that the strength of the violation can be substantially increased with respect to the non-optimized case (see Table \ref{tab_max_viol}). 
Note that, as expected, the violation strength diminishes for long chains and tends to the noninteracting $J=0$ curve, especially for short times. 

\begin{table}[hbtp]
%\vspace{2mm}
\center
\begin{tabular}{c| c c c c c c} 
 $n$ & 2 & 3 & 4 & 5 & 6 & 7\\
 \hline \hline
 NN  & 1.381 & 1.351 & 1.197 & 1.236 & 1.024 & 1.160\\
 \hline 
 NNN & 1.391 & 1.244 & 1.285 & 1.248 & 1.024 & 1.206\\
\end{tabular}
\caption{Maximal value of the $K_n$ correlator for times $ht\leq 1.8$ and the nearest neighbor Hamiltonian Eq.~\eqref{XXHamiltonian}  with $J=1, h = 1$ (NN) and its next--nearest-neighbor (NNN) extension.}
\label{tab_max_viol}
\end{table}

\begin{figure}[h]
  \centering
  \subfloat{\includegraphics[width=0.9\linewidth]{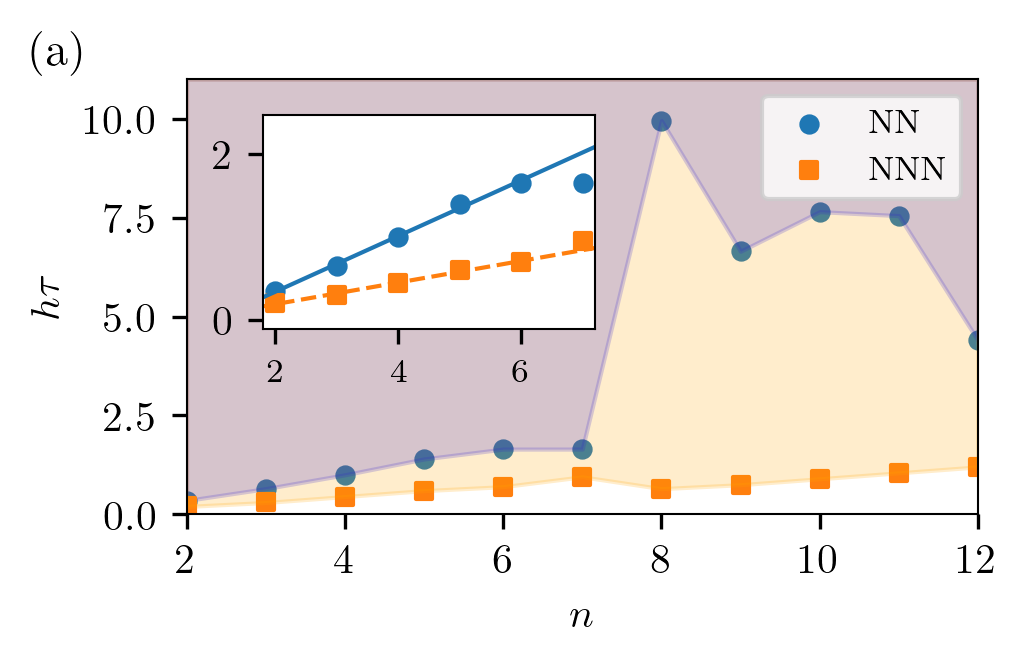}}
  
  \subfloat{\includegraphics[width=0.9\linewidth]{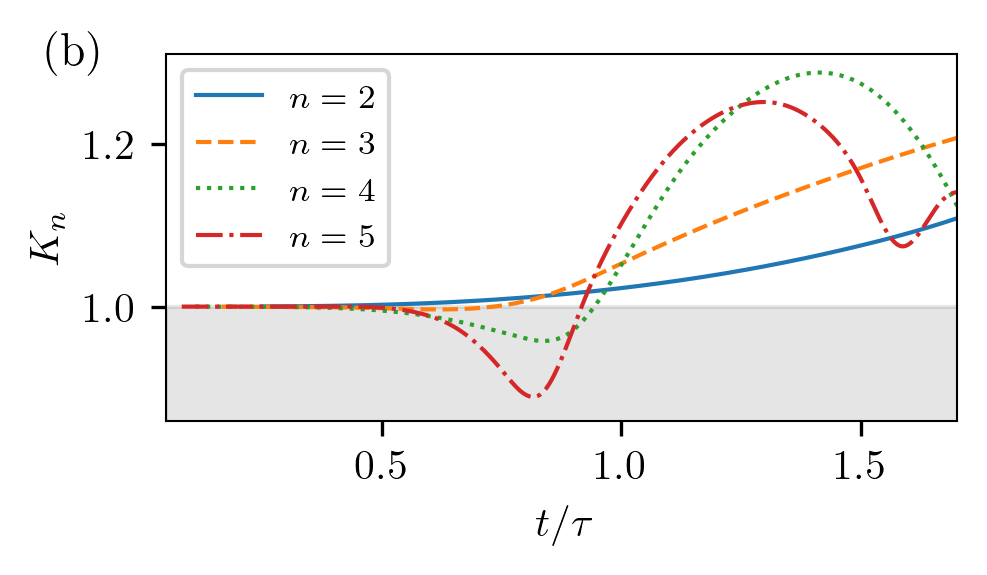}}
\caption{(a). Time of the first inequality violation $\tau$ versus distance $n$ (corresponding to $N=2n-1$ spins), showing that the speed of quantum correlations propagation in the spin chain is limited (we use the criterion $K_n>1.02$ to determine $\tau$). The points correspond to nearest-neighbor (NN) Hamiltonian Eq.~\eqref{XXHamiltonian} with $J=1$, $h = 1$ while the squares correspond to its next-nearest neighbor (NNN) extension (see text). The shaded regions exclude the Lieb Robinson time and, in the inset, the straight lines fit the points $2 \leq n \leq 6$ to simply emphasize their trend.
(b). Optimized spatial Leggett-Garg correlator versus normalized time for the NNN case and various distances $n$.
}
\label{fig3}
\end{figure}

We plot in Fig.~\ref{fig3}(a) the optimized first-violation times $h\tau$ versus $n$, which further demonstrates the relation between the spatial LGI violation and the correlation dynamics. 
In particular, to detect the violation and define $\tau$, we choose the conservative threshold $K_n \sim 1.02$ to be safely above the magnitude of the numerical errors and within the resolution of experiments \footnote{With ultracold atoms, correlators analogous to $C_{X,Y}$ are only measured with a few percent accuracy at most \cite{Cheneau2012}}. 
Our method provides a rigorous upper bound of the Lieb-Robinson time for \textit{each} system size $n$, or, equivalently, a lower bound on the Lieb-Robinson velocity. 
Note that, for nearest neighbor model data (NN) corresponding to $n\geq 8$, the first violation time quickly increases because such violation occurs in the second period of the corresponding non-interacting model (cf. Fig.~\ref{fig4}).
Moreover, for small distances $n$, we observe a linear behavior between $n$ and $h\tau$, see Fig.~\ref{fig3}(a) inset and Fig.~\ref{fig3}(b), which indicates that short-time correlations spread in a light-cone manner with constant velocity consistently with the Lieb-Robinson bounds. 

To further confirm the connection between Lieb-Robinson physics and the inequality violation, we include in Fig.~\ref{fig3}(a) next nearest-neighbor (NNN) interactions in the Hamiltonian.
In particular, by adding the term $J\sum_{i=1}^{N-2}\vec{\sigma}_i\cdot \vec{\sigma}_{i+2}$ to Eq.~\eqref{XXHamiltonian}, with $\vec{\sigma}_i = (\sigma^x_i, \sigma^y_i, \sigma^z_i)^T$, we observe again a linear proportionality between $\tau$ and $n$, but the slope of the curve decreases [orange squares in Fig.~\ref{fig3}(a)].
In other words, the spatial Leggett-Garg inequality violation is accelerated by enabling longer-range interactions that allow for a faster spreading of quantum correlations. 
Note that, for small system sizes, we actually obtain that the inequality is violated for any finite time $t>0$. 
This may occur because the sites $1,n,2n-1$ are directly coupled through the extended Hamiltonian for $n=2,3$, while for $n>3$ these sites do not interact directly and a finite time is needed to violate the LGI. \\

\begin{figure}
    \centering
    \includegraphics[width=\linewidth]{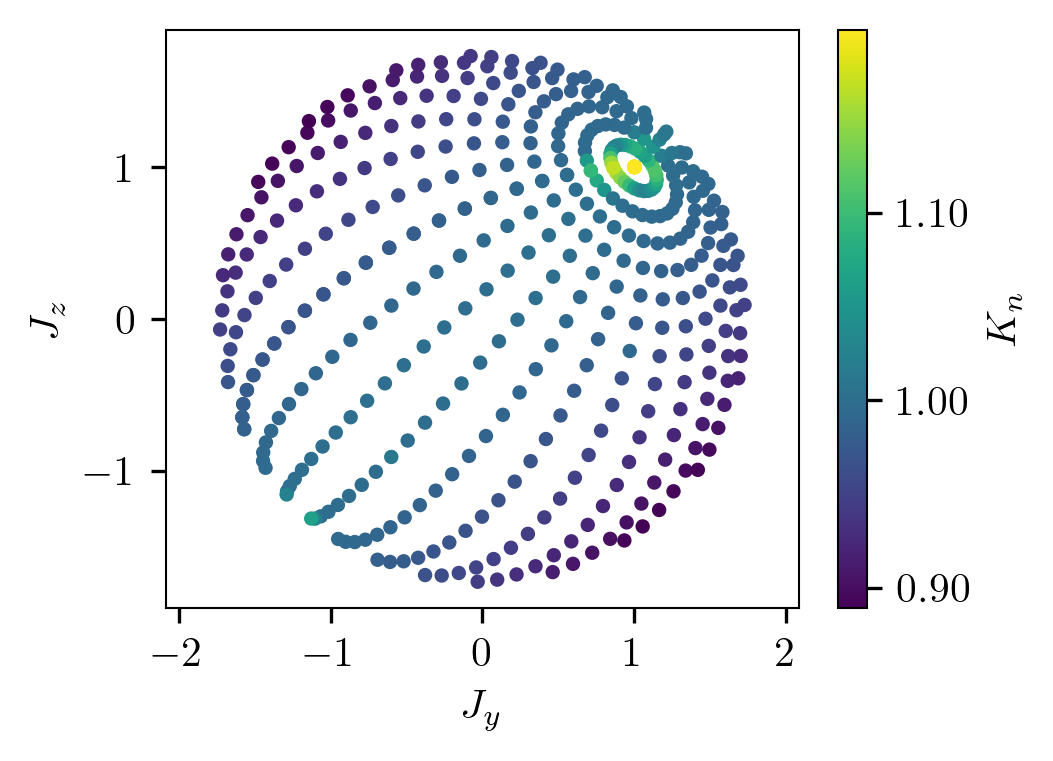}
    \caption{Maximal violation of the spatial Leggett-Garg inequality up to time $ht = 2$, evaluated for the Hamiltonian Eq.~\eqref{XXHamiltonian} with $n=4$ ($N = 7$).
Here we optimize over the measurements $Q$, take the initial state $\ket{+}^{\otimes N}$, and set $J_x =  \sqrt{3-J_y^2-J_z^2}$. 
The inequality violation is maximal for the Heisenberg model $J_x =J_y = J_z = 1$, under which correlations optimally spread.
Note that the results remain invariant under reflection of all couplings, i.e. under $J_{x,y,z} \to - J_{x,y,z}$.}
    \label{fig:th_phi}
\end{figure}

To gain further insight into which dynamics enable the violation of the spatial Leggett–Garg inequality, we consider the Hamiltonian~\eqref{XXHamiltonian} with the couplings constrained to $J_x^2 + J_y^2 +J_z^2 = 3$~\footnote{Since the inequality violation occurs after the Lieb-Robinson time, and the Lieb-Robinson velocity roughly scales with $J$ \cite{tononi2025}, this spherical constraint enables a fair comparison between the Hamiltonians \eqref{XXHamiltonian}.} and plot in Fig.~\ref{fig:th_phi} the maximal violation. We observe that the strongest violation occurs for isotropic interactions, while only marginal violation is observed for the XXZ model ($J_y = J_z$). 
Indeed, we note that the single-particle term (with coupling $h$) produces the maximal violation of the single-site LGI in this setting. 
The isotropic interacting term, in turn, corresponds to a chain of nearest-neighbor SWAP operators, and both terms commute. 
In the Heisenberg picture, the interacting term propagates operators along the chain so that, after sufficient time, they overlap. 
At that point, the single-particle term rotates the state, producing a notable violation in the multi-site LGI. 
The fact that isotropic interactions are optimal also facilitates the LGI implementation in the experimental platforms, where large anisotropic couplings are feasible, but nonetheless more challenging to achieve than isotropic ones~\cite{kim2024realization}.

Interestingly, the violation of the spatial LGI in our setting indicates that the Hamiltonian is close to a Heisenberg chain, which could be useful for Hamiltonian learning and certification protocols~\cite{Zhang2014, Wiebe2014}. Learning features of the Hamiltonian (for example, the couplings $J_x, J_y, J_z$) from experimental data alone is a highly resource intensive task and has been tackled with a plethora of methods ~\cite{Yu2023, Wilde2022, Gu2024}, also experimentally~\cite{Guo2025}. Our inequality, which is only based on three correlators and does not require trust in the measurement settings, may provide an advantage in this task.

The protocol of our work can be implemented in the available experimental platforms 
\cite{browaeys2020many, PRXQuantum.3.020303, geier2021floquet, miller2024two, bloch2012quantum, gross2017, doi:10.1126/science.abk2397, blatt2012quantum, RevModPhys.93.025001, jepsen2020spin, PhysRevX.11.011011}. 
For instance, ultracold atoms in optical lattices currently offer the possibility of preparing suitable initial states and performing a nonlocal time evolution with a prescribed Hamiltonian \cite{zhu2024}. By equipping such experiments with single-site addressability, the spin state can be imaged at the prescribed times and sites, and the measurement outcomes recorded. After enough rounds of collecting data, the statistics involved in LGI Eq.~\eqref{LG_spatial} (or in temporal Bell inequalities \cite{tononi2025}) can be inferred.

{Concerning the resilience of our results to environmental disturbance,
we envision
neutral atom devices \cite{doi:10.1126/science.abk2397}
 (e.g., quantum simulators) as suitable platforms for an experimental realization of our proposal. 
In these devices, especially optical tweezer arrays or Rydberg systems, the timescales $T_1$ and $T_2$ of lifetime and coherence respectively
can reach values on the order of seconds or longer \cite{saffman2016quantum, graham2022multi, covey2023} and provide inherent scalability.
As such,
engineering first violation times being $\ll T_{1,2}$ (hence a field $h$ of the order of $[0.1,10] \, s^{-1}$) would enable protection from these effects.
In the Appendix~\ref{sec:app_optimization}, we analyze the effect of perturbed measurements on the violation, showing that it remains robust. Nonetheless, an in-dept theoretical study in this direction would represent a compelling continuation of the present work. This could be accompanied by modeling platform-dependent effects of disorder and losses that could further reduce the measurement contrast. \\

\section{Conclusions}

In conclusion, we conceptually extend the Leggett-Garg inequality to analyze the correlation dynamics of a quantum many-body system. Our spatial Leggett–Garg inequality can be implemented on existing experimental platforms, providing a practical way to quantify their ability to induce spatiotemporal correlations. Moreover, it provides a rigorous lower bound on the Lieb-Robinson velocity in agnostic way, namely without full knowledge of the underlying Hamiltonian. Also, investigating the extension of these results to 2D and 3D systems could reveal insights about the spreading of quantum correlations in more complex geometries. {Finally, a natural direction for future research is to relate our construction to the twin problem of extending Bell inequalities to the temporal domain~\cite{Fritz2010, Dressel2014, tononi2025}, potentially through frameworks that aim to unify the description of space and time in non-relativistic quantum physics~\cite{Costa2018}. }\\

\noindent\textit{Acknowledgements}.---
We thank G. Rajchel-Mieldzioć and M. Płodzień for early discussions on the topic, and N. Gisin, A. Aspect and Č. Brukner for comments and discussions. 
G.M. and A.T. acknowledge financial support of the Horizon Europe programme HORIZON-CL4-2022-QUANTUM-02-SGA via Project No. 101113690 (PASQuanS2.1), and A. T. acknowledges funding by the European Union under the Horizon Europe MSCA programme via Project No. 101146753 (QUANTIFLAC).
D.F. acknowledges financial support from PNRR MUR Project No. PE0000023-NQSTI.
ICFO-QOT group acknowledges support from:
European Research Council AdG NOQIA; MCIN/AEI [Grants No. PGC2018-0910.13039/501100011033 and No. CEX2019-000910-S/10.13039/501100011033, Plan National STAMEENA PID2022-139099NB, project funded by MCIN and by the “European Union NextGenerationEU/PRTR" (PRTR-C17.I1), FPI]; QUANTERA DYNAMITE PCI2022-132919; QuantERA II Programme co-funded by European Union’s Horizon 2020 program under Grant Agreement No 101017733; Ministry for Digital Transformation and of Civil Service of the Spanish Government through the QUANTUM ENIA project call - Quantum Spain project, and by the European Union through the Recovery, Transformation and Resilience Plan - NextGenerationEU within the framework of the Digital Spain 2026 Agenda; Grant No. MICIU/AEI/10.13039/501100011033 and EU (No. PCI2025-163167); Fundació Cellex; 
Fundació Mir-Puig; Generalitat de Catalunya (European Social Fund FEDER and CERCA program; Barcelona Supercomputing Center MareNostrum (Grant No. FI-2023-3-0024); 
Funded by the European Union (Grants No. HORIZON-CL4-2022-QUANTUM-02-SGA, No. PASQuanS2.1, 101113690, EU Horizon 2020 FET-OPEN OPTOlogic, Grant No. 899794, QU-ATTO, Grant No. 101168628), EU Horizon Europe Program (Grant No.  101080086 NeQST and Grant Agreement No. 101080086 NeQST).
\\

\noindent\textit{Data availability statement}.--- The correlators are computed using exact diagonalization methods (for small $n$) and standard time-evolving block decimation (TEBD) and time dependent variational principle (TDVP) for the NN and NNN cases respectively implemented in TeNPy. Codes and data to generate the figures can be found in \cite{tobepublished}. 

\bibliography{bibfile}

@article{saffman2016quantum,
  title={Quantum computing with atomic qubits and {R}ydberg interactions: progress and challenges},
  author={Saffman, Mark},
  journal={J. Phys. B},
  volume={49},
  number={20},
  pages={202001},
  year={2016},
  publisher={IOP Publishing},
  doi={10.1088/0953-4075/49/20/202001}
}

@article{graham2022multi,
  title={Multi-qubit entanglement and algorithms on a neutral-atom quantum computer},
  author={Graham, T. M. and Song, Y. and Scott, J. and Poole, C. and Phuttitarn, L. and Jooya, K. and Eichler, P. and Jiang, X. and Marra, A. and Grinkemeyer, B. and others},
  journal={Nature},
  volume={604},
  number={7906},
  pages={457--462},
  year={2022},
  publisher={Nature Publishing Group UK London},
  doi={10.1038/s41586-022-04603-6}
}

@article{Fritz2010,
  title = {Quantum correlations in the temporal {C}lauser–{H}orne–{S}himony–{H}olt ({CHSH}) scenario},
  volume = {12},
  ISSN = {1367-2630},
  url = {http://dx.doi.org/10.1088/1367-2630/12/8/083055},
  DOI = {10.1088/1367-2630/12/8/083055},
  number = {8},
  journal = {New Journal of Physics},
  publisher = {IOP Publishing},
  author = {Fritz,  Tobias},
  year = {2010},
  month = aug,
  pages = {083055}
}

@article{vitagliano2023,
  title = {{L}eggett-{G}arg macrorealism and temporal correlations},
  author = {Vitagliano, Giuseppe and Budroni, Costantino},
  journal = {Phys. Rev. A},
  volume = {107},
  issue = {4},
  pages = {040101},
  numpages = {22},
  year = {2023},
  month = {Apr},
  publisher = {American Physical Society},
  doi = {10.1103/PhysRevA.107.040101},
  url = {https://link.aps.org/doi/10.1103/PhysRevA.107.040101}
}

@article{leggettgarg1985,
  title = {Quantum mechanics versus macroscopic realism: Is the flux there when nobody looks?},
  author = {{L}eggett, A. J. and {G}arg, Anupam},
  journal = {Phys. Rev. Lett.},
  volume = {54},
  issue = {9},
  pages = {857--860},
  numpages = {0},
  year = {1985},
  month = {Mar},
  publisher = {American Physical Society},
  doi = {10.1103/PhysRevLett.54.857},
  url = {https://link.aps.org/doi/10.1103/PhysRevLett.54.857}
}

@article{emary2014,
doi = {10.1088/0034-4885/77/1/016001},
url = {https://dx.doi.org/10.1088/0034-4885/77/1/016001},
year = {2013},
month = {dec},
publisher = {IOP Publishing},
volume = {77},
number = {1},
pages = {016001},
author = {Emary, Clive and Lambert, Neill and Nori, Franco},
title = {{L}eggett–{G}arg inequalities},
journal = {Reports on Progress in Physics}
}

@article{mendoza2019enhancing,
  title={Enhancing violations of {L}eggett-{G}arg inequalities in nonequilibrium correlated many-body systems by interactions and decoherence},
  author={Mendoza-Arenas, JJ and G{\'o}mez-Ruiz, FJ and Rodr{\'\i}guez, FJ and Quiroga, L},
  journal={Sci. Rep.},
  volume={9},
  number={1},
  pages={17772},
  year={2019},
  publisher={Nature Publishing Group UK London},
  doi={10.1038/s41598-019-54121-1}
}

@article{kofler2008,
  title = {Conditions for Quantum Violation of Macroscopic Realism},
  author = {Kofler, Johannes and Brukner, {\v{C}}aslav},
  journal = {Phys. Rev. Lett.},
  volume = {101},
  issue = {9},
  pages = {090403},
  numpages = {4},
  year = {2008},
  month = {Aug},
  publisher = {American Physical Society},
  doi = {10.1103/PhysRevLett.101.090403},
  url = {https://link.aps.org/doi/10.1103/PhysRevLett.101.090403}
}

@article{avis2010,
  title = {{L}eggett-{G}arg inequalities and the geometry of the cut polytope},
  author = {Avis, David and Hayden, Patrick and Wilde, Mark M.},
  journal = {Phys. Rev. A},
  volume = {82},
  issue = {3},
  pages = {030102},
  numpages = {4},
  year = {2010},
  month = {Sep},
  publisher = {American Physical Society},
  doi = {10.1103/PhysRevA.82.030102},
  url = {https://link.aps.org/doi/10.1103/PhysRevA.82.030102}
}

@article{lieb1972,
author = {Elliott H. Lieb and Derek W. Robinson},
title = {{The finite group velocity of quantum spin systems}},
volume = {28},
journal = {Communications in Mathematical Physics},
number = {3},
publisher = {Springer},
pages = {251 -- 257},
year = {1972},
doi={10.1007/BF01645779}
}

@article{wilde2012addressing,
  title={Addressing the clumsiness loophole in a {L}eggett-{G}arg test of macrorealism},
  author={Wilde, Mark M and Mizel, Ari},
  journal={Foundations of Physics},
  volume={42},
  pages={256--265},
  year={2012},
  publisher={Springer},
  doi={10.1007/s10701-011-9598-4}
}

@article{budroni2015quantum,
  title = {Quantum Nondemolition Measurement Enables Macroscopic {L}eggett-{G}arg Tests},
  author = {Budroni, C. and Vitagliano, G. and Colangelo, G. and Sewell, R. J. and G\"uhne, O. and T\'oth, G. and Mitchell, M. W.},
  journal = {Phys. Rev. Lett.},
  volume = {115},
  issue = {20},
  pages = {200403},
  numpages = {5},
  year = {2015},
  month = {Nov},
  publisher = {American Physical Society},
  doi = {10.1103/PhysRevLett.115.200403},
  url = {https://link.aps.org/doi/10.1103/PhysRevLett.115.200403}
}

@article{palacios2010experimental,
  title={Experimental violation of a {B}ell’s inequality in time with weak measurement},
  author={Palacios-Laloy, Agustin and Mallet, Fran{\c{c}}ois and Nguyen, Fran{\c{c}}ois and Bertet, Patrice and Vion, Denis and Esteve, Daniel and Korotkov, Alexander N},
  journal={Nat. Phys.},
  volume={6},
  number={6},
  pages={442--447},
  year={2010},
  publisher={Nature Publishing Group UK London},
  doi={10.1038/nphys1617}
}

@article{
doi:10.1126/science.abk2397,
author = {David Wei  and Antonio Rubio-Abadal  and Bingtian Ye  and Francisco Machado  and Jack Kemp  and Kritsana Srakaew  and Simon Hollerith  and Jun Rui  and Sarang Gopalakrishnan  and Norman Y. Yao  and Immanuel Bloch  and Johannes Zeiher },
title = {Quantum gas microscopy of Kardar-Parisi-Zhang superdiffusion},
journal = {Science},
volume = {376},
number = {6594},
pages = {716-720},
year = {2022},
doi = {10.1126/science.abk2397}
}

@article{browaeys2020many,
  title={Many-body physics with individually controlled {R}ydberg atoms},
  author={Browaeys, Antoine and Lahaye, Thierry},
  journal={Nat. Phys.},
  volume={16},
  number={2},
  pages={132--142},
  year={2020},
  publisher={Nature Publishing Group UK London},
  doi={10.1038/s41567-019-0733-z}
}

@article{jepsen2020spin,
  title={Spin transport in a tunable {H}eisenberg model realized with ultracold atoms},
  author={Jepsen, Paul Niklas and Amato-Grill, Jesse and Dimitrova, Ivana and Ho, Wen Wei and Demler, Eugene and Ketterle, Wolfgang},
  journal={Nature},
  volume={588},
  number={7838},
  pages={403--407},
  year={2020},
  publisher={Nature Publishing Group UK London},
  doi={10.1038/s41586-020-3033-y}
}

@article{PRXQuantum.3.020303,
  title = {Microwave Engineering of Programmable ${XX}Z$ {H}amiltonians in Arrays of {R}ydberg Atoms},
  author = {Scholl, P. and Williams, H. J. and Bornet, G. and Wallner, F. and Barredo, D. and Henriet, L. and Signoles, A. and Hainaut, C. and Franz, T. and Geier, S. and Tebben, A. and Salzinger, A. and Z\"urn, G. and Lahaye, T. and Weidem\"uller, M. and Browaeys, A.},
  journal = {PRX Quantum},
  volume = {3},
  issue = {2},
  pages = {020303},
  numpages = {10},
  year = {2022},
  month = {Apr},
  publisher = {American Physical Society},
  doi = {10.1103/PRXQuantum.3.020303}
}

@article{PhysRevX.11.011011,
  title = {Glassy Dynamics in a Disordered {H}eisenberg Quantum Spin System},
  author = {Signoles, A. and Franz, T. and Ferracini Alves, R. and G\"arttner, M. and Whitlock, S. and Z\"urn, G. and Weidem\"uller, M.},
  journal = {Phys. Rev. X},
  volume = {11},
  issue = {1},
  pages = {011011},
  numpages = {13},
  year = {2021},
  month = {Jan},
  publisher = {American Physical Society},
  doi = {10.1103/PhysRevX.11.011011}
}

@article{arnesen2001natural,
  title = {Natural Thermal and Magnetic Entanglement in the 1D {H}eisenberg Model},
  author = {Arnesen, M. C. and Bose, S. and Vedral, V.},
  journal = {Phys. Rev. Lett.},
  volume = {87},
  issue = {1},
  pages = {017901},
  numpages = {4},
  year = {2001},
  month = {Jun},
  publisher = {American Physical Society},
  doi = {10.1103/PhysRevLett.87.017901},
  url = {https://link.aps.org/doi/10.1103/PhysRevLett.87.017901}
}

@article{amico2008entanglement,
  title = {Entanglement in many-body systems},
  author = {Amico, Luigi and Fazio, Rosario and Osterloh, Andreas and Vedral, Vlatko},
  journal = {Rev. Mod. Phys.},
  volume = {80},
  issue = {2},
  pages = {517--576},
  numpages = {0},
  year = {2008},
  month = {May},
  publisher = {American Physical Society},
  doi = {10.1103/RevModPhys.80.517},
  url = {https://link.aps.org/doi/10.1103/RevModPhys.80.517}
}

@article{RevModPhys.93.025001,
  title = {Programmable quantum simulations of spin systems with trapped ions},
  author = {Monroe, C. and Camp{B}ell, W. C. and Duan, L.-M. and Gong, Z.-X. and Gorshkov, A. V. and Hess, P. W. and Islam, R. and Kim, K. and Linke, N. M. and Pagano, G. and Richerme, P. and Senko, C. and Yao, N. Y.},
  journal = {Rev. Mod. Phys.},
  volume = {93},
  issue = {2},
  pages = {025001},
  numpages = {57},
  year = {2021},
  month = {Apr},
  publisher = {American Physical Society},
  doi = {10.1103/RevModPhys.93.025001}
}

@article{
gross2017,
author = {Christian Gross  and Immanuel Bloch },
title = {Quantum simulations with ultracold atoms in optical lattices},
journal = {Science},
volume = {357},
number = {6355},
pages = {995-1001},
year = {2017},
doi = {10.1126/science.aal3837}
}

@article{bloch2012quantum,
  title={Quantum simulations with ultracold quantum gases},
  author={Bloch, Immanuel and Dalibard, Jean and Nascimbene, Sylvain},
  journal={Nat. Phys.},
  volume={8},
  number={4},
  pages={267--276},
  year={2012},
  publisher={Nature Publishing Group},
  doi={10.1038/nphys2259}
}

@article{blatt2012quantum,
  title={Quantum simulations with trapped ions},
  author={Blatt, Rainer and Roos, Christian F},
  journal={Nat. Phys.},
  volume={8},
  number={4},
  pages={277--284},
  year={2012},
  publisher={Nature Publishing Group UK London},
  doi={10.1038/nphys2252}
}

@article{geier2021floquet,
  title={{F}loquet {H}amiltonian engineering of an isolated many-body spin system},
  author={Geier, Sebastian and Thaicharoen, Nithiwadee and Hainaut, Cl{\'e}ment and Franz, Titus and Salzinger, Andre and Tebben, Annika and Grimshandl, David and Z{\"u}rn, Gerhard and Weidem{\"u}ller, Matthias},
  journal={Science},
  volume={374},
  number={6571},
  pages={1149--1152},
  year={2021},
  publisher={American Association for the Advancement of Science},
  doi={10.1126/science.abd9547}
}

@article{miller2024two,
  title={Two-axis twisting using {F}loquet-engineered XYZ spin models with polar molecules},
  author={Miller, Calder and Carroll, Annette N and Lin, Junyu and Hirzler, Henrik and Gao, Haoyang and Zhou, Hengyun and Lukin, Mikhail D and Ye, Jun},
  journal={Nature},
  volume={633},
  number={8029},
  pages={332--337},
  year={2024},
  publisher={Nature Publishing Group UK London},
  doi={10.1038/s41586-024-07883-2}
}

@article{budroni2013bounding,
  title = {Bounding Temporal Quantum Correlations},
  author = {Budroni, Costantino and Moroder, Tobias and Kleinmann, Matthias and G\"uhne, Otfried},
  journal = {Phys. Rev. Lett.},
  volume = {111},
  issue = {2},
  pages = {020403},
  numpages = {5},
  year = {2013},
  month = {Jul},
  publisher = {American Physical Society},
  doi = {10.1103/PhysRevLett.111.020403},
  url = {https://link.aps.org/doi/10.1103/PhysRevLett.111.020403}
}

@article{kim2024realization,
  title={Realization of an extremely anisotropic {H}eisenberg magnet in {R}ydberg atom arrays},
  author={Kim, Kangheun and Yang, Fan and M{\o}lmer, Klaus and Ahn, Jaewook},
  journal={Phys. Rev. X},
  volume={14},
  number={1},
  pages={011025},
  year={2024},
  publisher={APS},
  doi={10.1103/PhysRevX.14.011025}
}

@article{naimark-comment,
  title={},
  author={},
  journal={This aligns with Naimark's dilation theorem, according to which (applied to our case) a operator-valued measure (POVM) on a given spin can be implemented through a projective measurement on another ancillary spin that has previously interacted with it \cite{watrous2018theory}},
  volume={},
  number={},
  pages={},
  year={},
  publisher={}
}

@book{watrous2018theory,
  title={The theory of quantum information},
  author={Watrous, John},
  year={2018},
  publisher={Cambridge university press},
  doi={10.1017/9781316848142}
}

@misc{tobepublished,
  author       = {Guillem Müller-Rigat},
  title        = {{Spatial-{L}eggett-{G}arg}},
  year         = {2025},
  howpublished = {\url{https://github.com/GuillemMRR/Spatial-{L}eggett-{G}arg}},
  note         = {GitHub repository},
}

@article{Paz1993,
  title = {Proposed test for temporal {B}ell inequalities},
  author = {Paz, Juan Pablo and Mahler, G\"unter},
  journal = {Phys. Rev. Lett.},
  volume = {71},
  issue = {20},
  pages = {3235--3239},
  numpages = {0},
  year = {1993},
  month = {Nov},
  publisher = {American Physical Society},
  doi = {10.1103/PhysRevLett.71.3235},
  url = {https://link.aps.org/doi/10.1103/PhysRevLett.71.3235}
}

@article{GarcaMata2023,
  title = {Out-of-time-order correlations and quantum chaos},
  volume = {18},
  ISSN = {1941-6016},
  url = {http://dx.doi.org/10.4249/scholarpedia.55237},
  DOI = {10.4249/scholarpedia.55237},
  number = {4},
  journal = {Scholarpedia},
  publisher = {Scholarpedia},
  author = {Garça-Mata,  Ignacio and Jalabert,  Rodolfo and Wisniacki,  Diego},
  year = {2023},
  pages = {55237}
}

@article{Cubitt2008,
  title = {Engineering Correlation and Entanglement Dynamics in Spin Systems},
  author = {Cubitt, T. S. and Cirac, J. I.},
  journal = {Phys. Rev. Lett.},
  volume = {100},
  issue = {18},
  pages = {180406},
  numpages = {4},
  year = {2008},
  month = {May},
  publisher = {American Physical Society},
  doi = {10.1103/PhysRevLett.100.180406},
  url = {https://link.aps.org/doi/10.1103/PhysRevLett.100.180406}
}

@article{GmezRuiz2016,
  title = {Quantum phase transitions detected by a local probe using time correlations and violations of {L}eggett-{G}arg inequalities},
  author = {G\'omez-Ruiz, F. J. and Mendoza-Arenas, J. J. and Rodr\'{\i}guez, F. J. and Tejedor, C. and Quiroga, L.},
  journal = {Phys. Rev. B},
  volume = {93},
  issue = {3},
  pages = {035441},
  numpages = {10},
  year = {2016},
  month = {Jan},
  publisher = {American Physical Society},
  doi = {10.1103/PhysRevB.93.035441},
  url = {https://link.aps.org/doi/10.1103/PhysRevB.93.035441}
}

@article{lin2018,
  title = {Out-of-time-ordered correlators in a quantum Ising chain},
  author = {Lin, Cheng-Ju and Motrunich, Olexei I.},
  journal = {Phys. Rev. B},
  volume = {97},
  issue = {14},
  pages = {144304},
  numpages = {17},
  year = {2018},
  month = {Apr},
  publisher = {American Physical Society},
  doi = {10.1103/PhysRevB.97.144304},
  url = {https://link.aps.org/doi/10.1103/PhysRevB.97.144304}
}

@article{riddell2019,
  title = {Out-of-time ordered correlators and entanglement growth in the random-field {XX} spin chain},
  author = {Riddell, Jonathon and S\o{}rensen, Erik S.},
  journal = {Phys. Rev. B},
  volume = {99},
  issue = {5},
  pages = {054205},
  numpages = {13},
  year = {2019},
  month = {Feb},
  publisher = {American Physical Society},
  doi = {10.1103/PhysRevB.99.054205},
  url = {https://link.aps.org/doi/10.1103/PhysRevB.99.054205}
}

@article{tononi2025,
  title={Temporal {B}ell inequalities in non-relativistic many-body physics},
  author={Tononi, Andrea and Lewenstein, Maciej},
  journal={Quantum Sci. Technol.},
  volume={10},
  number={3},
  pages={03LT01},
  year={2025},
  publisher={IOP Publishing},
  doi={10.1088/2058-9565/adcbce}
}

@article{zhu2024,
  title={Splitting and connecting singlets in atomic quantum circuits},
  author={Zhu, Zijie and Kiefer, Yann and Jele, Samuel and G{\"a}chter, Marius and Bisson, Giacomo and Viebahn, Konrad and Esslinger, Tilman},
  journal={arXiv:2409.02984},
  year={2024},
  doi={10.48550/arXiv.2409.02984}
}

@article{santini2022,
  title = {Experimental violations of {L}eggett-{G}arg inequalities on a quantum computer},
  author = {Santini, Alessandro and Vitale, Vittorio},
  journal = {Phys. Rev. A},
  volume = {105},
  issue = {3},
  pages = {032610},
  numpages = {11},
  year = {2022},
  month = {Mar},
  publisher = {American Physical Society},
  doi = {10.1103/PhysRevA.105.032610},
  url = {https://link.aps.org/doi/10.1103/PhysRevA.105.032610}
}

@article{blasone2023,
  title = {{L}eggett-{G}arg inequalities in the quantum field theory of neutrino oscillations},
  author = {Blasone, Massimo and Illuminati, Fabrizio and Petruzziello, Luciano and Smaldone, Luca},
  journal = {Phys. Rev. A},
  volume = {108},
  issue = {3},
  pages = {032210},
  numpages = {7},
  year = {2023},
  month = {Sep},
  publisher = {American Physical Society},
  doi = {10.1103/PhysRevA.108.032210},
  url = {https://link.aps.org/doi/10.1103/PhysRevA.108.032210}
}

@article{naikoo2019,
  title = {{L}eggett-{G}arg inequality in the context of three flavor neutrino oscillation},
  author = {Naikoo, Javid and Alok, Ashutosh Kumar and Banerjee, Subhashish and Sankar, S. Uma},
  journal = {Phys. Rev. D},
  volume = {99},
  issue = {9},
  pages = {095001},
  numpages = {7},
  year = {2019},
  month = {May},
  publisher = {American Physical Society},
  doi = {10.1103/PhysRevD.99.095001},
  url = {https://link.aps.org/doi/10.1103/PhysRevD.99.095001}
}

@article{majidy2021,
  title = {Detecting violations of macrorealism when the original {L}eggett-{G}arg inequalities are satisfied},
  author = {Majidy, Shayan and Halliwell, Jonathan J. and Laflamme, Raymond},
  journal = {Phys. Rev. A},
  volume = {103},
  issue = {6},
  pages = {062212},
  numpages = {14},
  year = {2021},
  month = {Jun},
  publisher = {American Physical Society},
  doi = {10.1103/PhysRevA.103.062212},
  url = {https://link.aps.org/doi/10.1103/PhysRevA.103.062212}
}

@article{halliwell2017,
  title = {Comparing conditions for macrorealism: {L}eggett-{G}arg inequalities versus no-signaling in time},
  author = {Halliwell, J. J.},
  journal = {Phys. Rev. A},
  volume = {96},
  issue = {1},
  pages = {012121},
  numpages = {11},
  year = {2017},
  month = {Jul},
  publisher = {American Physical Society},
  doi = {10.1103/PhysRevA.96.012121},
  url = {https://link.aps.org/doi/10.1103/PhysRevA.96.012121}
}

@article{covey2023,
  title={Quantum networks with neutral atom processing nodes},
  author={Covey, Jacob P and Weinfurter, Harald and Bernien, Hannes},
  journal={npj Quantum Information},
  volume={9},
  number={1},
  pages={90},
  year={2023},
  publisher={Nature Publishing Group UK London},
  doi={10.1038/s41534-023-00759-9}
}

@article{Cheneau2012,
  title = {Light-cone-like spreading of correlations in a quantum many-body system},
  volume = {481},
  ISSN = {1476-4687},
  number = {7382},
  journal = {Nature},
  publisher = {Springer Science and Business Media LLC},
  author = {Cheneau,  Marc and Barmettler,  Peter and Poletti,  Dario and Endres,  Manuel and Schauß,  Peter and Fukuhara,  Takeshi and Gross,  Christian and Bloch,  Immanuel and Kollath,  Corinna and Kuhr,  Stefan},
  year = {2012},
  month = jan,
  pages = {484–487},
  DOI = {10.1038/nature10748},
  url = {http://dx.doi.org/10.1038/nature10748}
}

@article{Bonnes2014,
  title = {“Light-Cone” Dynamics After Quantum Quenches in Spin Chains},
  volume = {113},
  ISSN = {1079-7114},
  number = {18},
  journal = {Phys. Rev. Lett.},
  publisher = {American Physical Society (APS)},
  author = {Bonnes,  Lars and Essler,  Fabian H. L. and L\"{a}uchli,  Andreas M.},
  year = {2014},
  month = oct,
  DOI = {10.1103/physrevlett.113.187203}
}

@article{goggin2011,
author = {M. E. Goggin  and M. P. Almeida  and M. Barbieri  and B. P. Lanyon  and J. L. O’Brien  and A. G. White  and G. J. Pryde },
title = {Violation of the {L}eggett–{G}arg inequality with weak measurements of photons},
journal = {Proceedings of the National Academy of Sciences},
volume = {108},
number = {4},
pages = {1256-1261},
year = {2011},
doi = {10.1073/pnas.1005774108}}

@article{dressel2011,
  title = {Experimental Violation of Two-Party {L}eggett-{G}arg Inequalities with Semiweak Measurements},
  author = {Dressel, J. and Broadbent, C. J. and Howell, J. C. and Jordan, A. N.},
  journal = {Phys. Rev. Lett.},
  volume = {106},
  issue = {4},
  pages = {040402},
  numpages = {4},
  year = {2011},
  month = {Jan},
  publisher = {American Physical Society},
  doi = {10.1103/PhysRevLett.106.040402},
  url = {https://link.aps.org/doi/10.1103/PhysRevLett.106.040402}
}

@article{knee2012,
  title={Violation of a {L}eggett--{G}arg inequality with ideal non‐invasive measurements},
  author={Knee, George C and Simmons, Stephanie and Gauger, Erik M and Morton, John J.L. and Riemann, Helge and Abrosimov, Nikolai V and Becker, Peter and Pohl, Hans‐Joachim and Itoh, Kohei M and Thewalt, Mike L.W. and others},
  journal={Nat. Comm.},
  volume={3},
  number={1},
  pages={606},
  year={2012},
  publisher={Nature Publishing Group UK London},
  doi={10.1038/ncomms1614}
}

@article{athalye2011,
  title = {Investigation of the {L}eggett-{G}arg Inequality for Precessing Nuclear Spins},
  author = {Athalye, Vikram and Roy, Soumya Singha and Mahesh, T. S.},
  journal = {Phys. Rev. Lett.},
  volume = {107},
  issue = {13},
  pages = {130402},
  numpages = {5},
  year = {2011},
  month = {Sep},
  publisher = {American Physical Society},
  doi = {10.1103/PhysRevLett.107.130402},
  url = {https://link.aps.org/doi/10.1103/PhysRevLett.107.130402}
}

@article{zhou2015,
  title = {Experimental Detection of Quantum Coherent Evolution through the Violation of {L}eggett-{G}arg-Type Inequalities},
  author = {Zhou, Zong-Quan and Huelga, Susana F. and Li, Chuan-Feng and Guo, Guang-Can},
  journal = {Phys. Rev. Lett.},
  volume = {115},
  issue = {11},
  pages = {113002},
  numpages = {5},
  year = {2015},
  month = {Sep},
  publisher = {American Physical Society},
  doi = {10.1103/PhysRevLett.115.113002},
  url = {https://link.aps.org/doi/10.1103/PhysRevLett.115.113002}
}

@article{katiyar2017,
  title={Experimental violation of the {L}eggett--{G}arg inequality in a three‐level system},
  author={Katiyar, Hemant and Brodutch, Aharon and Lu, Dawei and Laflamme, Raymond},
  journal={New Journal of Physics},
  volume={19},
  number={2},
  pages={023033},
  year={2017},
  publisher={IOP Publishing},
  doi={10.1088/1367-2630/19/023033}
}

@article{wang2020,
  title = {Experimental violations of {L}eggett-{G}arg inequalities up to the algebraic maximum for a photonic qubit},
  author = {Wang, Kunkun and Xu, Mengyan and Xiao, Lei and Xue, Peng},
  journal = {Phys. Rev. A},
  volume = {102},
  issue = {2},
  pages = {022214},
  numpages = {5},
  year = {2020},
  month = {Aug},
  publisher = {American Physical Society},
  doi = {10.1103/PhysRevA.102.022214},
  url = {https://link.aps.org/doi/10.1103/PhysRevA.102.022214}
}

@article{zhan2023,
  title = {Experimental violation of the {L}eggett-{G}arg inequality in a three-level trapped-ion system},
  author = {Zhan, Tianxiang and Wu, Chunwang and Zhang, Manchao and Qin, Qingqing and Yang, Xueying and Hu, Han and Su, Wenbo and Zhang, Jie and Chen, Ting and Xie, Yi and Wu, Wei and Chen, Pingxing},
  journal = {Phys. Rev. A},
  volume = {107},
  issue = {1},
  pages = {012424},
  numpages = {9},
  year = {2023},
  month = {Jan},
  publisher = {American Physical Society},
  doi = {10.1103/PhysRevA.107.012424},
  url = {https://link.aps.org/doi/10.1103/PhysRevA.107.012424}
}

@article{liu2019,
  title = {Strict experimental test of macroscopic realism in a light-matter-interfaced system},
  author = {Liu, Xiao and Zhou, Zong-Quan and Han, Yong-Jian and Li, Zong-Feng and Hu, Jun and Yang, Tian-Shu and Li, Pei-Yun and Liu, Chao and Li, Xue and Ma, Yu and Liang, Peng-Jun and Li, Chuan-Feng and Guo, Guang-Can},
  journal = {Phys. Rev. A},
  volume = {100},
  issue = {3},
  pages = {032106},
  numpages = {8},
  year = {2019},
  month = {Sep},
  publisher = {American Physical Society},
  doi = {10.1103/PhysRevA.100.032106},
  url = {https://link.aps.org/doi/10.1103/PhysRevA.100.032106}
}

@article{knee2016,
  title={A strict experimental test of macroscopic realism in a superconducting flux qubit},
  author={Knee, George C and Kakuyanagi, Kosuke and Yeh, Mao-Chuang and Matsuzaki, Yuichiro and Toida, Hiraku and Yamaguchi, Hiroshi and Saito, Shiro and {L}eggett, Anthony J and Munro, William J},
  journal={Nat. Comm.},
  volume={7},
  number={1},
  pages={13253},
  year={2016},
  publisher={Nature Publishing Group UK London},
  doi={10.1038/ncomms13253}
}

@article{wang2017,
  title={Enhanced violations of {L}eggett‐{G}arg inequalities in an experimental three‐level system},
  author={Wang, Kunkun and Emary, Clive and Zhan, Xiang and Bian, Zhihao and Li, Jian and Xue, Peng},
  journal={Optics Express},
  volume={25},
  number={25},
  pages={31462--31470},
  year={2017},
  publisher={Optical Society of America},
  doi={10.1364/OE.25.031462}
}

@article{formaggio2016,
  title = {Violation of the {L}eggett-{G}arg Inequality in Neutrino Oscillations},
  author = {Formaggio, J. A. and Kaiser, D. I. and Murskyj, M. M. and Weiss, T. E.},
  journal = {Phys. Rev. Lett.},
  volume = {117},
  issue = {5},
  pages = {050402},
  numpages = {6},
  year = {2016},
  month = {Jul},
  publisher = {American Physical Society},
  doi = {10.1103/PhysRevLett.117.050402},
  url = {https://link.aps.org/doi/10.1103/PhysRevLett.117.050402}
}

@article{kreuzgruber2024,
  title = {Violation of a {L}eggett-{G}arg Inequality Using Ideal Negative Measurements in Neutron Interferometry},
  author = {Kreuzgruber, Elisabeth and Wagner, Richard and Geerits, Niels and Lemmel, Hartmut and Sponar, Stephan},
  journal = {Phys. Rev. Lett.},
  volume = {132},
  issue = {26},
  pages = {260201},
  numpages = {6},
  year = {2024},
  month = {Jun},
  publisher = {American Physical Society},
  doi = {10.1103/PhysRevLett.132.260201},
  url = {https://link.aps.org/doi/10.1103/PhysRevLett.132.260201}
}

@article{Chatterjee2025,
  title = {Extreme Violations of {L}eggett-{G}arg Inequalities for a System Evolving under Superposition of Unitaries},
  volume = {135},
  ISSN = {1079-7114},
  url = {http://dx.doi.org/10.1103/vydp-9qqq},
  DOI = {10.1103/vydp-9qqq},
  number = {22},
  journal = {Physical Review Letters},
  publisher = {American Physical Society (APS)},
  author = {Chatterjee,  Arijit and Karthik,  H. S. and Mahesh,  T. S. and Usha Devi,  A. R.},
  year = {2025},
  month = nov 
}

@article{Lu2024,
  title = {Experimental demonstration of enhanced violations of {L}eggett-{G}arg inequalities in asymmetric trapped-ion qubit},
  volume = {109},
  ISSN = {2469-9934},
  url = {http://dx.doi.org/10.1103/PhysRevA.109.042205},
  DOI = {10.1103/physreva.109.042205},
  number = {4},
  journal = {Physical Review A},
  publisher = {American Physical Society (APS)},
  author = {Lu,  Pengfei and Rao,  Xinxin and Liu,  Teng and Liu,  Yang and Bian,  Ji and Zhu,  Feng and Luo,  Le},
  year = {2024},
  month = apr 
}

@article{Cheng2025,
  title = {{B}ell Inequality Violation of Light Quarks in Dihadron Pair Production at Lepton Colliders},
  volume = {135},
  ISSN = {1079-7114},
  url = {http://dx.doi.org/10.1103/gmqz-v4cl},
  DOI = {10.1103/gmqz-v4cl},
  number = {1},
  journal = {Physical Review Letters},
  publisher = {American Physical Society (APS)},
  author = {Cheng,  Kun and Yan,  Bin},
  year = {2025},
  month = jul 
}

@article{Zhang2014,
  title = {Quantum {H}amiltonian Identification from Measurement Time Traces},
  volume = {113},
  ISSN = {1079-7114},
  url = {http://dx.doi.org/10.1103/PhysRevLett.113.080401},
  DOI = {10.1103/physrevlett.113.080401},
  number = {8},
  journal = {Physical Review Letters},
  publisher = {American Physical Society (APS)},
  author = {Zhang,  Jun and Sarovar,  Mohan},
  year = {2014},
  month = aug 
}

@article{Guo2025,
  title = {{H}amiltonian learning for 300 trapped ion qubits with long-range couplings},
  volume = {11},
  ISSN = {2375-2548},
  url = {http://dx.doi.org/10.1126/sciadv.adt4713},
  DOI = {10.1126/sciadv.adt4713},
  number = {5},
  journal = {Science Advances},
  publisher = {American Association for the Advancement of Science (AAAS)},
  author = {Guo,  Shi-An and Wu,  Yu-Kai and Ye,  Jing and Zhang,  Lin and Wang,  Ye and Lian,  Wen-Qian and Yao,  Rui and Xu,  Yu-Lin and Zhang,  Chi and Xu,  Yu-Zi and Qi,  Bin-Xiang and Hou,  Pan-Yu and He,  Li and Zhou,  Zi-Chao and Duan,  Lu-Ming},
  year = {2025},
  month = jan 
}

@article{Yu2023,
  title = {Robust and Efficient {H}amiltonian Learning},
  volume = {7},
  ISSN = {2521-327X},
  url = {http://dx.doi.org/10.22331/q-2023-06-29-1045},
  DOI = {10.22331/q-2023-06-29-1045},
  journal = {Quantum},
  publisher = {Verein zur Forderung des Open Access Publizierens in den Quantenwissenschaften},
  author = {Yu,  Wenjun and Sun,  Jinzhao and Han,  Zeyao and Yuan,  Xiao},
  year = {2023},
  month = jun,
  pages = {1045}
}

@article{Wiebe2014,
  title = {{H}amiltonian Learning and Certification Using Quantum Resources},
  volume = {112},
  ISSN = {1079-7114},
  url = {http://dx.doi.org/10.1103/PhysRevLett.112.190501},
  DOI = {10.1103/physrevlett.112.190501},
  number = {19},
  journal = {Physical Review Letters},
  publisher = {American Physical Society (APS)},
  author = {Wiebe,  Nathan and Granade,  Christopher and Ferrie,  Christopher and Cory,  D. G.},
  year = {2014},
  month = may 
}

@misc{Wilde2022,
  doi = {10.48550/ARXIV.2209.14328},
  url = {https://arxiv.org/abs/2209.14328},
  author = {Wilde,  Frederik and Kshetrimayum,  Augustine and Roth,  Ingo and Hangleiter,  Dominik and Sweke,  Ryan and Eisert,  Jens},
  title = {Scalably learning quantum many-body {H}amiltonians from dynamical data},
  publisher = {arXiv},
  year = {2022},
  copyright = {Creative Commons Attribution 4.0 International}
}

@article{Gu2024,
  title = {Practical {H}amiltonian learning with unitary dynamics and {G}ibbs states},
  volume = {15},
  ISSN = {2041-1723},
  url = {http://dx.doi.org/10.1038/s41467-023-44008-1},
  DOI = {10.1038/s41467-023-44008-1},
  number = {1},
  journal = {Nature Communications},
  publisher = {Springer Science and Business Media LLC},
  author = {Gu,  Andi and Cincio,  Lukasz and Coles,  Patrick J.},
  year = {2024},
  month = jan 
}

@article{Dressel2014,
  title = {Avoiding loopholes with hybrid {B}ell-{L}eggett-{G}arg inequalities},
  volume = {89},
  ISSN = {1094-1622},
  url = {http://dx.doi.org/10.1103/PhysRevA.89.012125},
  DOI = {10.1103/physreva.89.012125},
  number = {1},
  journal = {Physical Review A},
  publisher = {American Physical Society (APS)},
  author = {Dressel,  Justin and Korotkov,  Alexander N.},
  year = {2014},
  month = Jan 
}

@article{Costa2018,
  title = {Unifying framework for spatial and temporal quantum correlations},
  volume = {98},
  ISSN = {2469-9934},
  url = {http://dx.doi.org/10.1103/PhysRevA.98.012328},
  DOI = {10.1103/physreva.98.012328},
  number = {1},
  journal = {Physical Review A},
  publisher = {American Physical Society (APS)},
  author = {Costa,  Fabio and Ringbauer,  Martin and Goggin,  Michael E. and White,  Andrew G. and Fedrizzi,  Alessandro},
  year = {2018},
  month = July 
}

\pagebreak
\appendix

\section{Noninteracting Hamiltonians}
\label{sec:app_noniteracting}
Here we discuss the spatial LGI for noninteracting Hamiltonians, i.e. for those containing only local couplings.
To begin with, we consider the noninteracting version of Eq.~\eqref{XXHamiltonian}, i.e. $H = -(h/2)\sum_i \sigma^z_i$, and derive the dynamics of $K_n$ for the measurement $Q = \sigma^x $ and the state $\ket{\Psi}$. 
Elementary calculations show that $C_{1,n} = \cos(ht)$,  $C_{n,2n} = \cos(ht)\cos(2ht) $, and $C_{1,2n-1} = \cos(2ht) $ which add up to $K_n = \left[ 3 \cos(h t) - 2 \cos(2h t) + \cos(3h t)  \right]/2$. 
This quantity, depicted as the black solid line of Fig.~\ref{fig4}, does not violate the LGI.
The result of Fig.~\ref{fig4} prompts us to consider interactions to change the $J=0$ curve and observe violation of the spatial Leggett-Garg inequality (cf. Fig.~\ref{fig5} in this appendix). 
Indeed, interactions couple the spins which dynamically become non separable.
Roughly speaking, time evolution brings the state of the spin chain closer to the one of a single quantum system, which violates the inequality (see, for instance, the blue dashed curve in Fig.~\ref{fig4}) \cite{watrous2018theory,naimark-comment}.

\begin{figure}[hbtp]
\centering
\includegraphics[width=0.99\linewidth]{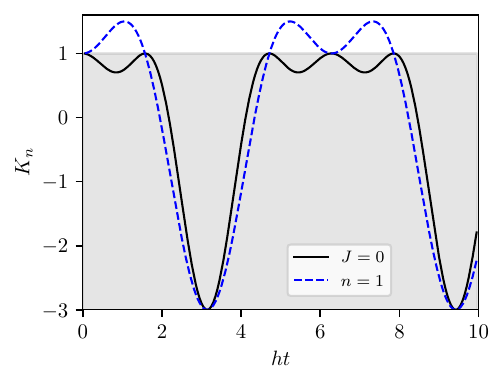}
\caption{Comparison of the Leggett-Garg dynamics for a single spin $n=1$ and its spatial extension for non-interacting Hamiltonian $J = 0$ for any $n>1$ and observable $Q = \sigma^x$.
}
\label{fig4}
\end{figure}

More generally, we can prove that, under evolution governed by noninteracting Hamilonians, the spatial LGI cannot be violated. Indeed, for noninteracting evolution, the observables $(Q_1(0), Q_n(t),Q_{2n-1}(2t))$ commute pairwise, as it happens for $t=0$. Thus, they are jointly measurable with outcomes $(q_1, q_n, q_{2n-1}) \in \{\pm 1 \}$ with probability $p(q_1,q_n, q_{2n-1})$ (which depends on $\ket{\Psi}$). Thus the value of the correlator is $K_n= \sum_{q_1, q_n, q_{2n-1}} p(q_1, q_n, q_{2n-1}) (q_1q_n + q_nq_{2n-1} - q_{1}q_{2n-1})$. By convexity, the maximal of $K_N$ is attained by a deterministic strategy $p(q_1, q_n, q_{2n-1}) = \delta_{q_1,a}\delta_{q_2,b}\delta_{q_3,c}$, with $\delta $ the Kronecker's delta. Simple enumeration of all the values of $K_n$ for $(a,b,c)\in \{\pm 1\}$ verifies that $K_n$ is confined between $-3$ and $1$. Note that the claim holds regardless the initial state $\ket{\Psi}$ (separable or entangled) and the details of the local measurement $Q$.

\section{Optimization over measurements} 
\label{sec:app_optimization}
In this appendix, we detail the procedure to optimize the correlator over measurements. Specifically, we find the measurement orientation $\vec{v}\in \mathbb{S}^2$, $\sigma^v = \vec{v}\cdot \vec{\sigma}$, such that $K_n$ is maximal. To that end, we evaluate $K_n$ for $Q = \sigma^v$, obtaining $K_n(\vec{v}) = \vec{v}^T \mathcal{K}_n \vec{v}$, where $\mathcal{K}_n$ is a $3\times 3$ symmetric matrix with elements $[\mathcal{K}]_{pq} = \Re{\bra{\Psi}\hat{\mathcal{K}}_{pq}  \ket{\Psi}}$, with $p,q\in \{ x,y,z\}\ $ and $\hat{\mathcal{K}}_{pq} = \sigma_1^{p}(0)\sigma_n^{q}(n) + \sigma_n^{p}(n)\sigma_{2n-1}^{q}(2n) - \sigma_1^{p}(0)\sigma_{2n-1}^{q}(2n)$. From the bilinear structure of the correlator, it is direct to see that  $K_n(\vec{v}) \leq  \lambda_{\max}(\mathcal{K}_n) = K_n(\vec{v}_{\max})$, where $\lambda_{\max}$ denotes the maximal eigenvalue and $\vec{v}_{\max}$ the corresponding eigenvector (the optimal measurement is $\sigma^{\rm opt} = \vec{v}_{\max}\cdot \vec{\sigma}$). In Fig.~\ref{fig5} we display an example of such optimization for distance $n = 5$ in the presence of noise disturbing the measurement settings. To that end, we perturb the ideal setting with Gaussian noise $Q\propto Q^{\rm ideal} + \epsilon \mathbf{n}\cdot \boldsymbol{\sigma}$, where $\mathbf{n} = (n^x, n^y, n^z)$ are samples dependent of the normal distribution. We display the average and fluctuations of the dynamics of the Leggett-Garg correlator over various samples. 

\begin{figure}[h!]
\centering
\includegraphics[width=0.99\linewidth]{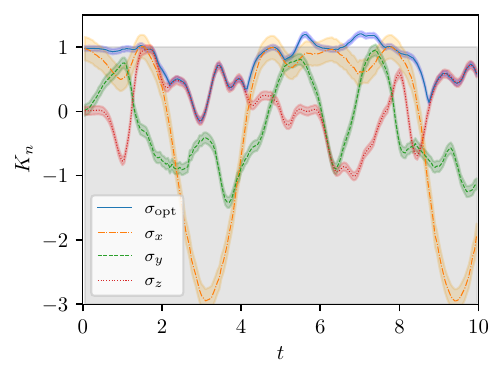}
\caption{
Average dynamics of the Leggett-Garg correlator for measurement settings $Q$ for $Q^{\rm ideal}  = \sigma^x, \sigma^y, \sigma^z$ and noise level $\epsilon = 0.1$ for $n=5$. We also display the data for the best setting $Q^{\rm ideal} = \sigma^{\rm opt}$ optimized at each time instance and subject to the same noise level $\epsilon = 0.1$.}
\label{fig5}
\end{figure}

Note that from an experimental perspective, the evaluation of the maximal nonlocality requires the inference of the matrix $\mathcal{K}$ involving sequential expectation values of the form $\langle \sigma^p_i(t_i) \sigma^q_j(t_j) \rangle$ for all spin components $p,q \in \{x,y,z\}$. We also observe that the violation is robust against noise and that different observables are affected differently; for example, $\sigma_x$ is more sensitive to noise than $\sigma_y$ and $\sigma_z$. We leave the optimization of the measurements, taking noise into account, for future studies. \\

Next, in Tab.~\ref{tab_max_viol}, we specify the maximal violation for short times, which suggests that extended Hamiltonians not only accelerate the onset of violation but also strengthen its magnitude. 

\newpage

\end{document}